\documentclass{arxiv}
\usepackage[T1]{fontenc}
\usepackage[latin9]{inputenc}
\usepackage{amstext}
\usepackage{graphicx}

\title{Study of X-ray Radiation Damage in Silicon Sensors}

\author{Jiaguo Zhang$^a$\thanks{Corresponding
author.}~, Eckhart Fretwurst$^a$, Robert Klanner$^a$, Hanno Perrey$^a$, Ioana Pintilie$^b$, Thomas Poehlsen$^a$, and Joern Schwandt$^a$\\
\llap{$^a$}Institute for Experimental Physics, Hamburg University,\\
  Luruper Chaussee 149, D-22761 Hamburg, Germany\\
\llap{$^b$}National Institute of Materials Physics,\\
  P.O.Box MG-7, Bucharest-Magurele, Romania\\
  E-mail: \email{jiaguo.zhang@desy.de}}

\abstract{The European X-ray Free Electron Laser (XFEL) will deliver 30,000
fully coherent, high brilliance X-ray pulses per second each with a duration
below 100 fs. This will allow the recording of diffraction patterns of single complex
molecules and the study of ultra-fast processes. Silicon pixel sensors will
be used to record the diffraction images. In 3 years of operation
the sensors will be exposed to doses of up to 1 GGy of 12 keV X-rays.
At this X-ray energy no bulk damage in silicon is expected. However
fixed oxide charges in the insulating layer covering the silicon and interface traps at the Si-SiO$_{\text{2}}$ interface will be introduced by the irradiation and build up over time.
\\

We have investigated the microscopic
defects in test structures and the macroscopic electrical properties
of segmented detectors as a function of the X-ray dose. From the test structures we determine the oxide
charge density and the densities of interface traps as a function of dose.
We find that both saturate (and even decrease) for doses between 10 and 100
MGy. For segmented sensors the defects introduced by the X-rays increase
the full depletion voltage, the surface leakage current and the inter-pixel
capacitance. We observe that an electron accumulation layer forms at the
Si-SiO$_{\text{2}}$ interface. Its width increases with dose and decreases
with applied bias voltage. Using TCAD simulations with the dose dependent parameters
obtained from the test structures, we are able to reproduce the observed
results. This allows us to optimize the sensor design for the XFEL
requirements.
\\

In addition the Si-SiO$_{\text{2}}$ interface region has been studied with
time resolved signals induced by sub-nanosecond 660 nm laser light, which has a penetration
of about 3 $\mu$m in silicon. Depending on the biasing history, humidity
and irradiation dose, losses of either electrons or holes or no charge
losses are observed. The relevance of these results for the sensor stability
and performance is under investigation.}

\keywords{XFEL; pixel sensors; surface radiation damage; simulations; charge losses}

\begin{document}


\section{Introduction}

Imaging experiments at the European X-ray Free Electron Laser (XFEL), planned to be operational in 2014, require silicon pixel sensors with extraordinary performance specifications: A dynamic range from single to 10$^{\text{5}}$ 12 keV photons deposited in less than 100 fs in one pixel of 200 $\mu$m $\times$ 200  $\mu$m, a time interval between XFEL pulses of 220 ns, and radiation doses up to 1 GGy for 3 years of operation. To optimize the sensor performance for these requirements, especially for radiation tolerance, demands an excellent understanding of the radiation damage caused by X-rays and its effects on segmented sensors.

The maximum energy transfer to silicon atoms from 12 keV X-rays is 0.011 eV, which is far below the threshold energy of 21 eV for bulk damage \cite{bib6} and therefore no bulk damage is expected. X-rays produce electron-hole pairs in the insulating layer (typically SiO$_{\text{2}}$ or SiO$_{\text{2}}$/Si$_{\text{3}}$N$_{\text{4}}$). Some of these charge carriers produced by the 12 keV X-rays recombine, whereas those remaining either remain in the insulating layer or move to the electrode on top of the insulating layer or drift to the Si-SiO$_{\text{2}}$ interface. Once holes come close to the Si-SiO$_{\text{2}}$ interface a fraction of them will be trapped in the oxide close to the interface, and produce radiation induced fixed oxide charges. In addition, border traps \cite{bib7} and interface traps are produced: The former are located in the oxide near the Si-SiO$_{\text{2}}$ interface and can be charged up or discharged through capture and emission of electrons and/or holes; whereas the latter are at the Si-SiO$_{\text{2}}$ interface and have energy levels distributed throughout the silicon band gap and whose occupation therefore changes with band bending. They are responsible for the surface generation current. The densities of fixed oxide charges and interface traps introduced by X-rays mainly depend on the dose (energy deposition in the insulating layer), dose rate, electric field during irradiation, and post-irradiation conditions (e.g. time and temperature of annealing).

The fixed oxide charges and interface traps at the Si-SiO$_{\text{2}}$ interface change the performance of segmented silicon sensors: This manifests itself as a shift in the full depletion voltage to larger values, an increase of the leakage current and the inter-pixel capacitance, a decrease of the inter-pixel resistance, and formation of an electron accumulation layer, which may cause charge losses at the interface. In this work we have irradiated MOS capacitors to determine the microscopic radiation damage parameters, and characterized the electrical properties of p$^+$n silicon microstrip sensors irradiated to doses up to 100 MGy.

\section{Irradiation with X-rays}

\subsection{Irradiation facility}
For the irradiations, a facility has been set up \cite{bib1} at the beam line F4 of HASYLAB DORIS III, DESY. The set-up is shown in figure 1. It consists of an adjustable Ta chopper, which permits the reduction of the full dose rate (180 kGy/s) down to 0.5\%, an adjustable collimator to precisely define the region of irradiation, and a sample holder with 5 biasing lines, which is connected to a liquid cooling system to control the temperature in the range of 10 to 30 $^{\circ}$C.

The test structures and sensors were glued and wire-bonded to a ceramic substrate, shown at the top left of figure 1. This holder makes the exchange and test of the structures easy and safe.

The set-up is mounted on a table, which permits computer control of both the horizontal and vertical movement of the sample enabling larger areas than the beam spot to be irradiated uniformly. A pneumatic beam shutter, for a precise determination of the exposure time to the beam, was installed close to the exit window of the beam pipe.

\begin{figure}[htbp]
\small
\centering
\includegraphics[scale=0.5]{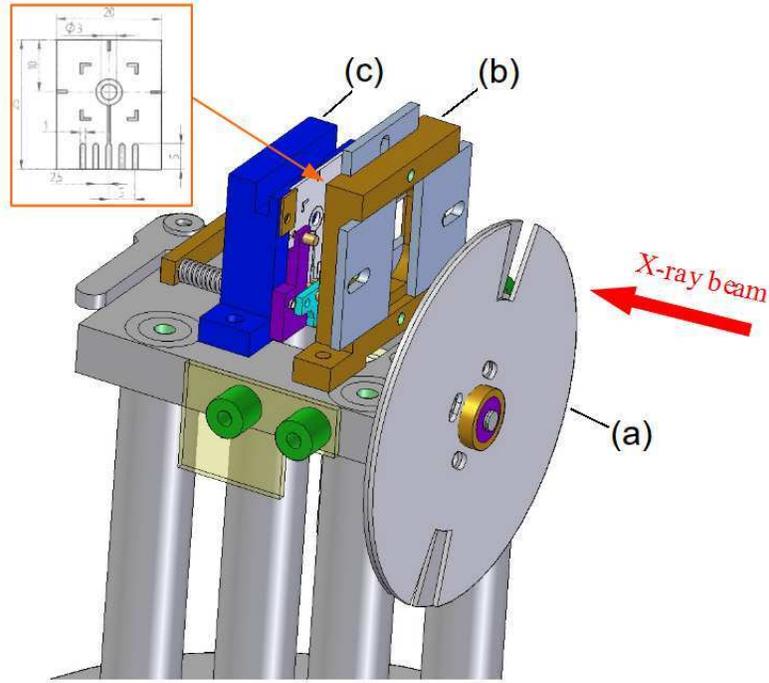}
\caption{Set-up of the irradiation facility: Chopper (a), collimator (b), and sample holder (c) with cooling and spring mechanism. The ceramic substrate which is put into the sample holder carries test structures and sensors (shown at the top left). The entire set-up can be moved in all three dimensions by computer control.}
\label{Figure1}
\end{figure}

\subsection{Beam profile}

The beam line F4 at DORIS III provides a "white" photon beam from a bending magnet. The energy spectrum was calculated taking into account the X-ray absorption by the materials in the beam line. As shown in figure 2(a) the typical photon energy at the detector was 12 keV with the full width at half maximum of about 10 keV. 

The profile of the X-ray beam was measured by the photocurrent in a silicon pad diode biased to 6 V. The pad diode, glued onto a ceramic substrate as shown in figure 1, was placed onto the sample holder. The collimator was opened to a window of 0.2 mm $\times$ 0.2 mm. The sample was moved in steps of 0.1 mm to scan the beam. The photon flux as a function of position was calculated from the measured current, the average energy of 3.6 eV to produce one electron-hole pair in silicon and the energy deposited by the photons. The fraction of interacting photons was derived from the energy spectrum and the absorption length of X-rays as a function of energy. Figure 2(b) shows the measured two dimensional beam profile.

The horizontal and vertical profiles at the center of the beam are shown in figure 2(c) and figure 2(d). The width of the horizontal profile is about 5 mm and uniform; whereas the full width of half maximum of the vertical profile is 4 mm and Gaussian-like. The distributions are as expected for the synchrotron radiation of the positrons of DORIS III in a bending magnet.

\begin{figure}[htbp]
\small
\centering
\includegraphics[scale=0.5]{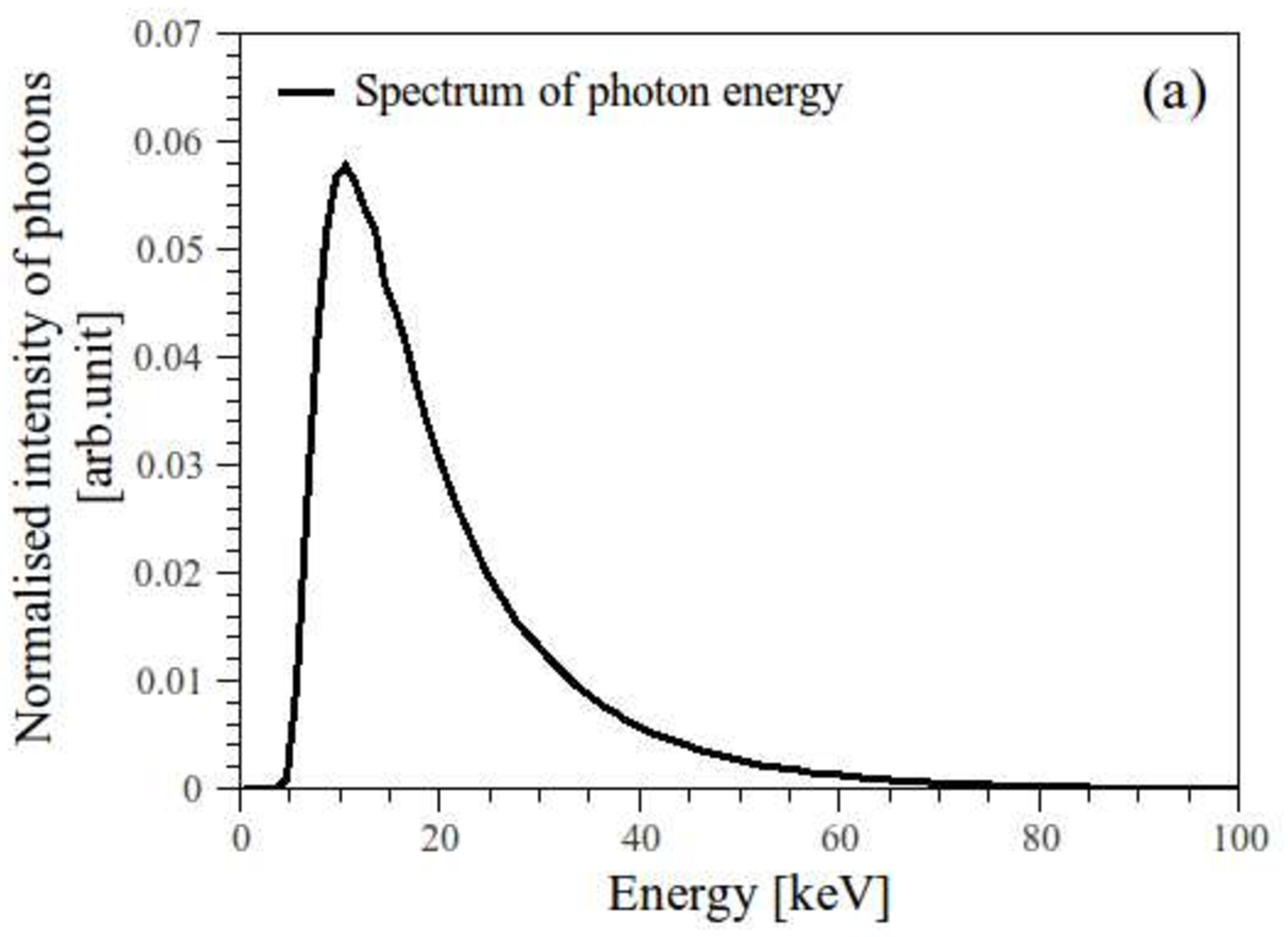}
\includegraphics[scale=0.34]{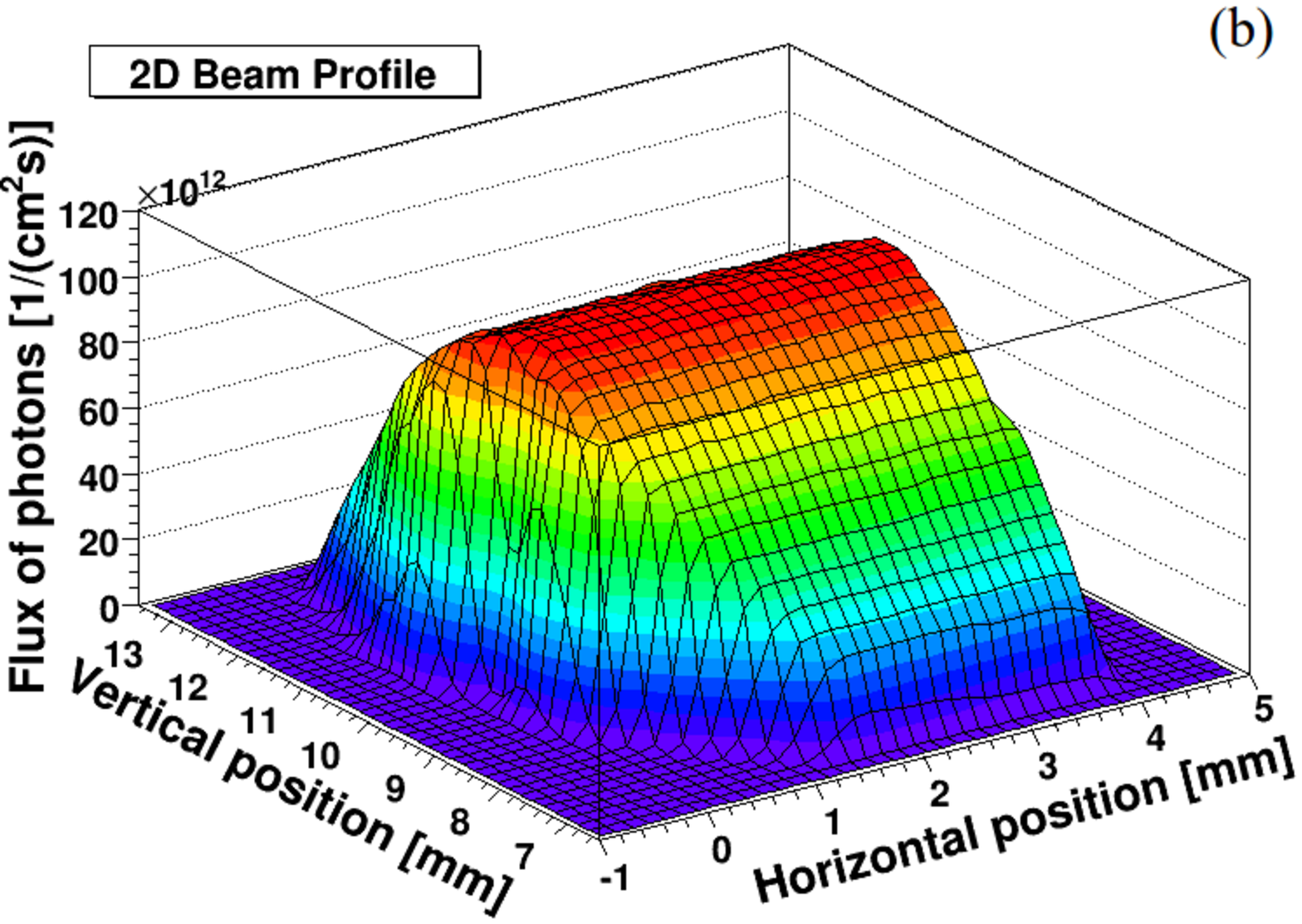}
\includegraphics[scale=0.5]{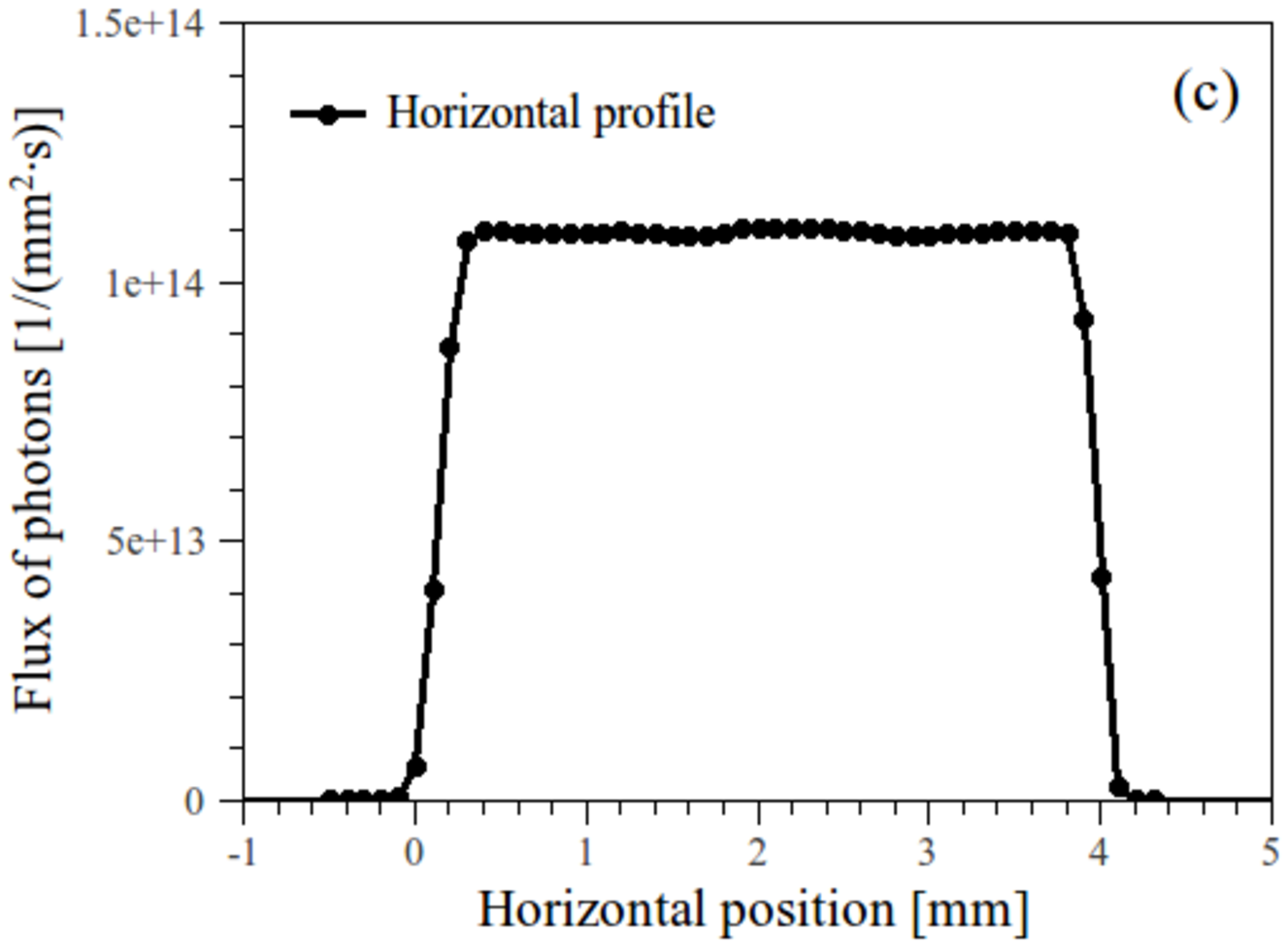}
\includegraphics[scale=0.5]{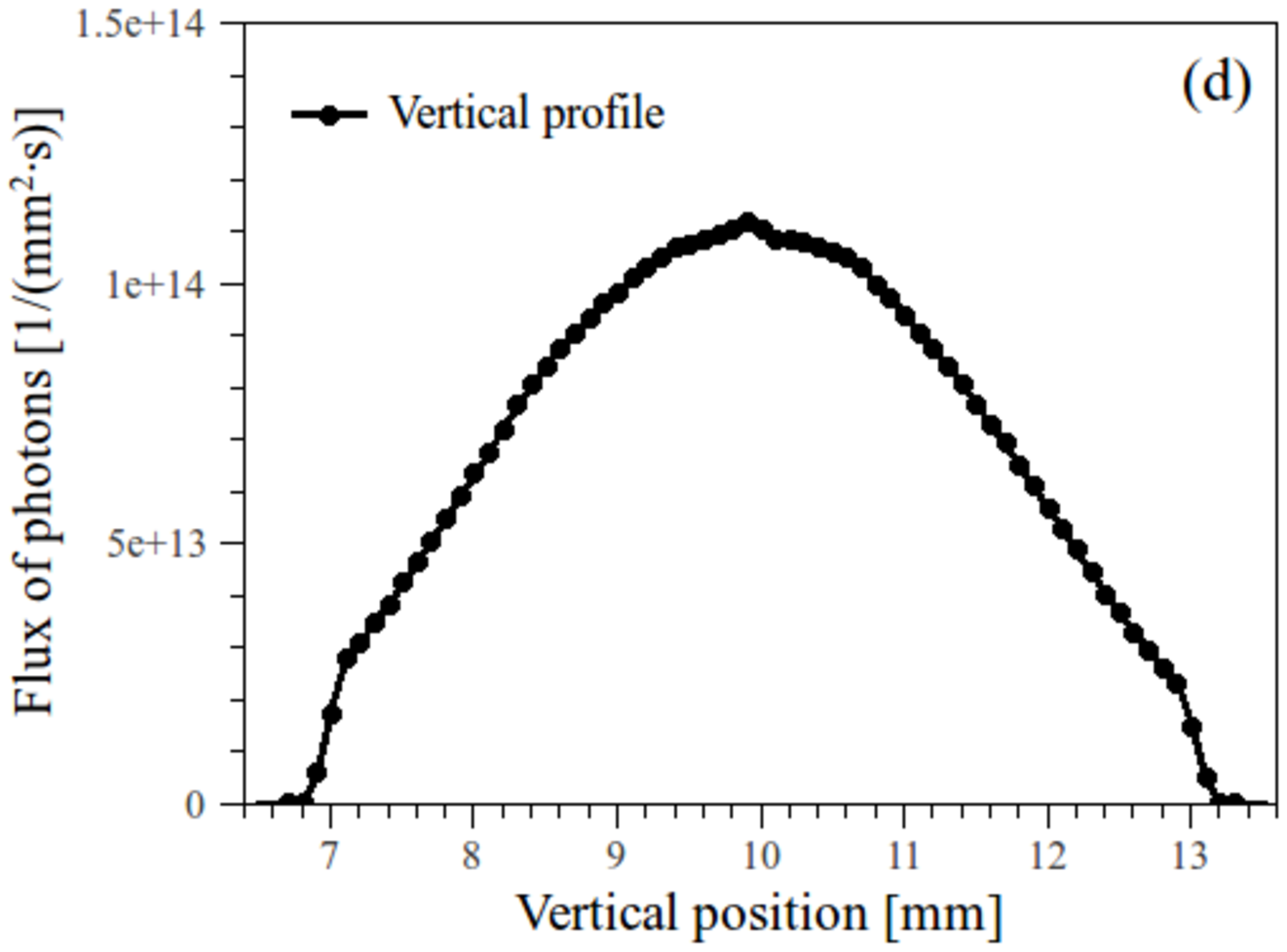}
\caption{(a) Normalised spectrum of photon energy. (b) Two dimensional beam profile. (c) Horizontal profile. (d) Vertical profile.}
\label{Figure2}
\end{figure}

\subsection{Determination of the dose}

Using the data shown in figure 2(a) the normalised photon spectrum and figure 2(b) the two dimensional beam profile, the average dose in the insulator can be calculated according to:

\begin{equation}
\label{eq:dose}
	Dose =  \frac{3.6eV \cdot R}{q_{0}\rho _{Si}d_{Si}A}
\int\!\!\!\!\int\!\!\!\int_{\Delta x, \Delta y, t_{irra}} J_{diode}(x, y) \cdot 
dxdydt
\end{equation}

with $J_{diode}(x, y)$ the current density ($A/mm^{2}$) measured by the silicon diode, $t_{irra}$ the duration of irradiation, $\rho_{Si}$ and $d_{Si}$ the density and thickness of the silicon diode. $A$ is the irradiated area of the detector, $\Delta x$ and $\Delta y$ the horizontal and vertical extensions of the detector in the plane perpendicular to the beam and $q_{0}$ is the elementary charge. $R$ is the ratio of energy deposited per unit mass in the insulator ($E_{Insulator}$) to energy deposited in the silicon  ($E_{Si}$) for the given photon spectrum.

\begin{equation}
\label{eq:rate}
	R = \frac{\frac{E_{Insulator}}{\rho_{Insulator} \cdot d_{Insulator}}}{\frac{E_{Si}}{\rho_{Si} \cdot d_{Si}}}
\end{equation}

$\rho_{Insulator}$ and  $d_{Insulator}$ are the density and thickness of the insulator. For the given energy spectrum, and $d_{SiO_{2}}$ = 300 nm and $d_{Si}$ = 300 $\mu$m, the value of $R$ is 1.30, which is essentially independent of $d_{SiO_{2}}$ for the typical oxide thickness of sensors. Without the chopper the typical dose rates in SiO$_{2}$ are 180 kGy/s, which corresponds to a flux of approximately 1.1$\times$10$^{14}$ photons/(mm$^{2}$ $\cdot$s).

The results from the measurement of the dose agree within 20\% with the calculations using the current of the DORIS III beam, the field of the bending magnet and the geometry of the set-up. All references to dose in the text refer to the surface dose in SiO$_{2}$. The dose enhancement in SiO$_{2}$ has not been taken into account in this study \cite{bib10}.

\section{Extraction of the concentrations of defects induced by X-rays from MOS capacitors}

To study the radiation induced damage at the Si-SiO$_{\text{2}}$ interface, the Metal-Oxide-Semiconductor (MOS) capacitor was used. The MOS capacitors under study were fabricated by CiS on 280 $\mu$m thick n-doped <100> substrates with a doping concentration of about 7.0$\times$10$^{\text{11}}$ cm$^{-3}$. The insulator is made of 350 nm SiO$_{\text{2}}$ covered by 50 nm Si$_{\text{3}}$N$_{\text{4}}$. The diameter of the \textasciitilde 1 $\mu$m thick circular aluminium gate is 1.5 mm.

The MOS capacitors were irradiated to SiO$_{2}$ doses from 12 kGy up to 1 GGy. The dose rate was 18 kGy/s, except for the irradiation to 1 GGy (180 kGy/s) and to 12 kGy (1.3 kGy/s). The results obtained do not strongly depend on the dose rate according to our previous investigations \cite{bib1}.

\subsection{Method}

In a previous study of one highly irradiated MOS capacitor, it was found that one of the Si-SiO$_{2}$ interface traps showed a fast annealing already at room temperature which made the measurements unreproducible. Therefore, in order to obtain reproducible results, all irradiated MOS capacitors were annealed for 10 minutes at 80 $^\circ$C before the measurements.

\subsubsection{Experimental techniques}

The Capacitance-Voltage (C-V) and Conductance-Voltage (G-V) measurements have been performed at room temperature using an Agilent E4980A bridge. An AC voltage of 50 mV and frequencies from 1 kHz to 1 MHz have been chosen. The voltage scan started at 0 V (strong accumulation in our case) and stopped before strong inversion was reached in order to avoid injecting holes into border traps \cite{bib4}.

In order to determine the parameters and the densities of interface traps the Thermally Dielectric Relaxation Current (TDRC) technique \cite{bib2, bib9} was employed. The TDRC technique is a well-established tool to study majority carrier traps at the Si-SiO$_{\text{2}}$ interface of MOS capacitors. The TDRC measurement was carried out as follow: The MOS capacitor was cooled to low temperature (10 K using a helium cooling system) biased in strong accumulation to fill the interface traps with majority carriers (with electrons for n-type silicon). At 10 K the gate voltage was changed to bias the MOS capacitor in weak inversion, in which condition traps remained filled since the temperature was too low to allow the thermal emission of the electrons. Then the sample was heated up to room temperature with a constant rate of $\beta$ = 0.183 K/s. As the temperature rises the trapped electrons are gradually released through thermal emission and the trap discharge current is recorded as TDRC signal. In general, the TDRC signal is due to both Si-SiO$_{\text{2}}$ interface traps and traps in the depletion region of the bulk silicon. However, since the 12 keV X-rays only introduce interface traps, the TDRC signal from bulk traps can be ignored. 

Figure 3 shows the TDRC signal as a function of temperature for different doses after annealing at 80 $^\circ$C for 10 minutes. A pronounced peak at \textasciitilde 225 K, corresponding to an energy of 0.60 eV as measured from the conduction band, and a broad distribution between 210 K and 50 K can be observed. The TDRC signal increases with dose up to 100 MGy and decreases at 1 GGy. The decrease of the TDRC signal is probably due to the annealing of traps at high dose rates during the longer irradiation time.

\begin{figure}[htbp]
\small
\centering
\includegraphics[scale=0.8]{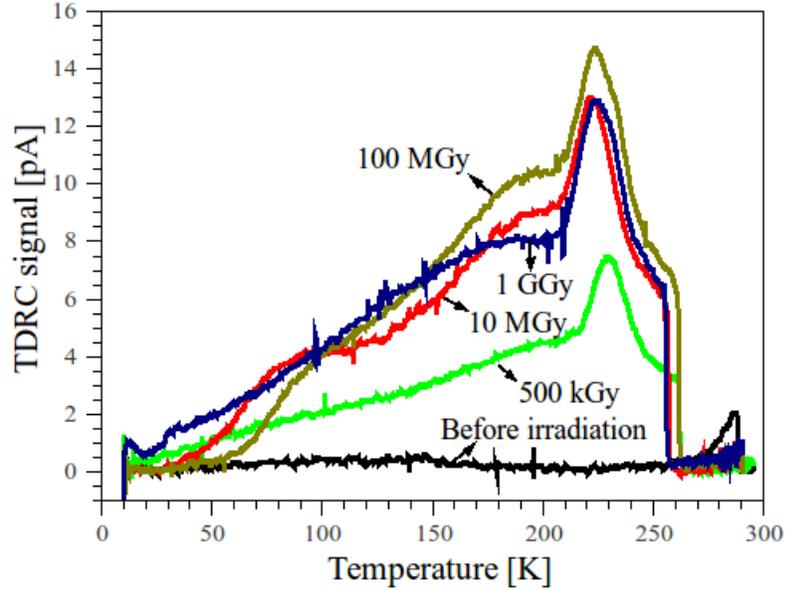}
\caption{TDRC spectra ($\beta$ = 0.183 K/s) of the MOS capacitors for different irradiations after annealing for 10 minutes at  80 $^\circ$C. Only 0 kGy (before irradiation), 500 kGy, 10 MGy, 100 MGy and 1GGy are shown.}
\label{Figure3}
\end{figure}

In the analysis, we assumed that the interface states density of each trap are uniformly distributed in space with Gaussian-like energy distributions in the silicon band gap. Three Gaussian distributions were used to describe the measured TDRC spectra. The energy distribution and the density of interface states of each trap were calculated using the equations (16) and (17) in the paper of H. A. Mar \cite{bib3} for separated, Gaussian distributed TDRC signals. For this calculation, electron capture cross sections for each trap are needed, which were obtained by minimizing the $\chi^{\text{2}}$  of the difference between the measured and calculated C/G values for the frequencies 1, 3, 10 and 30 kHz at different gate voltages. The model calculation is discussed in the next section. Parameters, e.g. electron capture cross section, mean energy and full width of half maximum of each trap have been reported in \cite{bib4}.


\subsubsection{Model calculation}

The following model which includes the effects of the interface traps was used to describe the C/G-V measurements of the MOS capacitors. As shown in figure 4, the model consists of an RC circuit. The equations to calculate the capacitance and the resistance of each element in the circuit were described in \cite{bib4}. The admittance of the circuit as a function of gate voltage was calculated based on the measured TDRC spectra, evaluated electron capture cross sections, doping concentration and fixed oxide charge density, and finally compared to the measured parallel capacitance and conductance. The doping concentration of the semiconductor close to the Si-SiO$_{2}$ interface was determined from the high frequency capacitance in strong inversion following the Lindner approximation \cite{bib5}. The three interface traps were assumed to be acceptors. The fixed oxide charge density $N_{\text{ox}}$, which just causes a shift in voltage, were extracted from the voltage shifts observed in the measured C/G-V curves.

\begin{figure}[htbp]
\small
\centering
\includegraphics[scale=0.5]{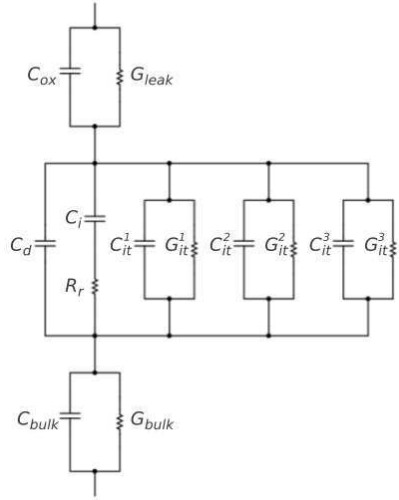}
\caption{Model with three interface traps used to calculate the C/G-V curves for the MOS capacitors: $C_{\text{ox}}$ and $G_{\text{leak}}$ is the capacitance and conductance of the insulator; $C_{\text{d}}$, the capacitance of the depletion layer; $C_{\text{i}}$, the inversion capacitance due to minority carriers accumulating at the interface; $R_{\text{r}}$, the recombination/generation resistance; $C_{\text{it}}^{\text{i}}$ and $G_{\text{it}}^{\text{i}}$, the capacitances and conductances due to interface traps; $C_{\text{bulk}}$ and $G_{\text{bulk}}$, the capacitance and conductance of the non-depleted silicon bulk. The relations between $C_{\text{it}}^{\text{i}}$ and $G_{\text{it}}^{\text{i}}$ and the microscopic properties of the traps are given in \cite{bib4}.}
\label{Figure4}
\end{figure}


\subsection{Results}

It was found \cite{bib4} that, after annealing at 80 $^{\circ}$C for 10 minutes, the fixed oxide charge density $N_{ox}$ and the interface trap densities $N_{it}^{1,2,3} = \int D_{it}^{1,2,3}(E_{t}) dE_{t}$ integrated over the band gap saturate at doses between 10 and 100 MGy. The saturation value of $N_{ox}$ and $N_{it}^{1,2,3}$ are 2.8$\times$10$^{12}$, 1.2$\times$10$^{12}$, 0.7$\times$10$^{12}$ and 0.6$\times$10$^{12}$ cm$^{-2}$, respectively. The extracted microscopic parameters related to the surface radiation damage caused by the X-rays have been implemented in Synopsys TCAD simulation \cite{bib8} with the aim of optimizing the sensor design for the XFEL applications.


\section{Influence on the performance of segmented sensors}

To experimentally study the influence of the radiation induced damage on the performance of segmented sensors, we use AC-coupled p$^+$n microstrip sensors. The microstrip sensors were fabricated by CiS on an n-doped silicon substrate with a thickness of 280 $\mu$m and with the orientation <100>.  The doping concentration in bulk silicon is about 8.0$\times$10$^{\text{11}}$ cm$^{\text{-3}}$ (as determined from the C-V curve of a close-by pad diode on the same wafer). The AC coupling between the p$^+$ implantation and the aluminium strips is made of 200 nm SiO$_{\text{2}}$ and 50 nm Si$_{\text{3}}$N$_{\text{4}}$. In between the aluminium strips the silicon is covered by 300 nm SiO$_{\text{2}}$ and 50 nm Si$_{\text{3}}$N$_{\text{4}}$. The gap between neighbouring implants is 62 $\mu$m, the pitch 80 $\mu$m and the length of the 98 strips of each sensor 7.9 mm. All implanted strips are connected to a biasing ring through biasing resistors, which are made with a low dose p$^+$ implantation. The biasing ring, the current collection ring, namely the innermost ring close to the biasing ring, and 13 guard rings surround all strips and define the active detection area of the sensor.

The p$^+$n microstrip sensors mounted onto the ceramic substrates as discussed in section 2.1, were irradiated to SiO$_{2}$ doses of 1 MGy, 10 MGy and 100 MGy. After irradiation, in order to obtain reproducible results, the sensors were annealed at 80 $^{\circ}$C for 60 minutes before the measurements. After the 100 MGy irradiation the biasing resistors failed. As shown in figure 5(a), the biasing resistance, derived from the slope of the I-V measurement between one of the strips and the biasing ring, exceeds 100 M$\Omega$ after irradiation to 100 MGy. Whereas the values of the biasing resistor before irradiation, and after 1 and 10 MGy are 0.5, 0.6 and 1.0 M$\Omega$. The failure of the biasing resistor can be explained by the removal of free holes in the low dose p$^+$ implants due to the positive oxide and interface charges. In the following we therefore will only present the results obtained for dose values of 0, 1 and 10 MGy. 

\subsection{Electrical properties}

The Capacitance-Voltage (C-V) and  Current-Voltage (I-V) measurements have been performed at room temperature. For the measurement of the capacitance of the entire sensor, a DC power supply was connected to the rear side of the sensor through a 1 k$\Omega$ resistor; the high frequency terminal of the Agilent 4980A was connected to the rear side of the sensor through a capacitor of 1 $\mu$F and the AC current from the biasing ring of the sensor fed back to the input terminal of the bridge. The current collection ring was grounded and the guard rings were kept floating during the measurement. Figure 5(b) shows the capacitance of the entire sensor measured at 100 kHz,  plotted as 1/C$^{2}$ versus bias voltage. 1/C$^{2}$ increases linearly with bias voltage and saturates at the full depletion voltage V$_{\text{dep}}$. A kink at \textasciitilde 6 V, the voltage at which the depletion regions below the individual strips merge, is observed. The full depletion capacitance C$_{\text{dep}}$ expected for a pad diode with the same area and thickness of the silicon substrate is 23.3 pF, close to the measured value of 23.6 pF for the microstrip sensors. It should be noted that, compared to a pad diode, an additional voltage is needed for microstrip sensors to fully deplete the sensor. The full depletion voltage V$_{\text{dep}}$ of the microstrip sensor before irradiation is about 12 V larger than the value calculated from the same doping concentration and thickness for a pad diode. After irradiations, the full depletion voltage V$_{\text{dep}}$ increases by an additional \textasciitilde 10 V, which is required to compensate for the positive charges in the oxide and at the Si-SiO$_{2}$ interface. The increase of the full depletion voltage V$_{\text{dep}}$ saturates at high doses. Furthermore, the radiation introduced interface traps can be charged or discharged by an external AC signal, and thus act like a frequency dependent capacitor as seen in figure 5(c). This effect is implemented in the model for the MOS capacitor discussed in section 3.1.2.

The leakage current was measured from the biasing ring of the sensor by a Keithley 6517A electrometer. The biasing ring and the rear side of the sensor were connected to the current input terminal and the voltage output terminal of the electrometer, respectively. A Keithley 6485 picoammeter was used to monitor the current flowing into the current collection ring. All guard rings were left floating. The results of the measurements are shown in figure 5(d). Compared to the I-V curve without irradiation, the leakage currents after irradiations increase by two orders of magnitude and do not show any saturation with bias even above the full depletion voltage V$_{\text{dep}}$. The leakage current is mainly due to the surface generation current from the Si-SiO$_{2}$ interface regions not covered by an electron accumulation layer. The observed linear increase of the leakage current with bias voltage agrees with the results of the Synopsys TCAD simulation, which shows that the accumulation layer decreases approximately linearly with bias voltage. The measured leakage current agrees with the surface current of \textasciitilde 8 $\mu$A/cm$^{2}$ measured for highly irradiated gate controlled diodes \cite{bib1}. 


Interstrip capacitance and interstrip resistance (not shown here), which are directly related to the electron accumulation layer at the Si-SiO$_{2}$ interface, increase with dose and decrease with applied bias voltage, as expected from the dose and voltage dependence of the width of the accumulation layer.

\begin{figure}[htbp]
\small
\centering
\includegraphics[scale=0.49]{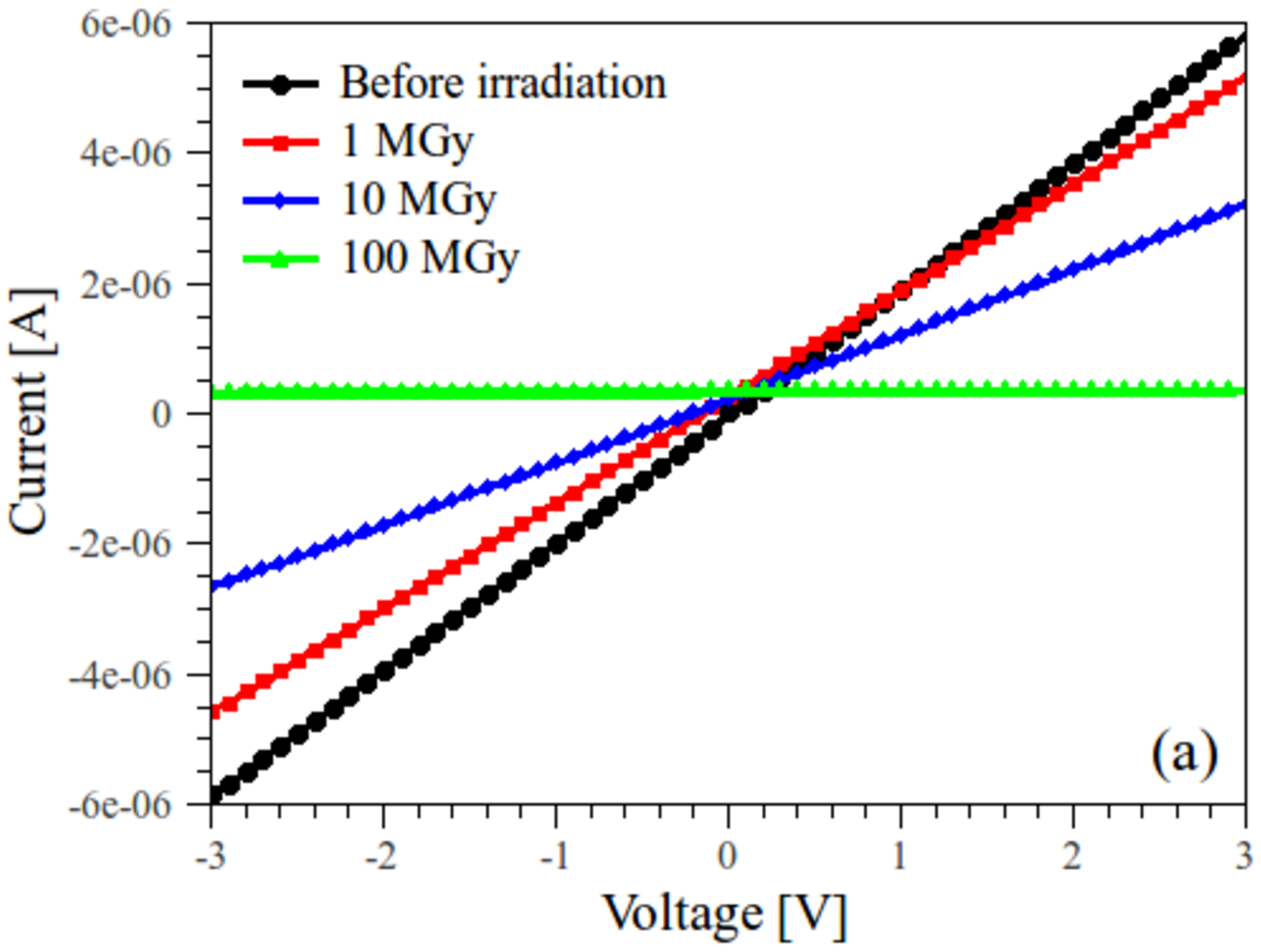}
\includegraphics[scale=0.49]{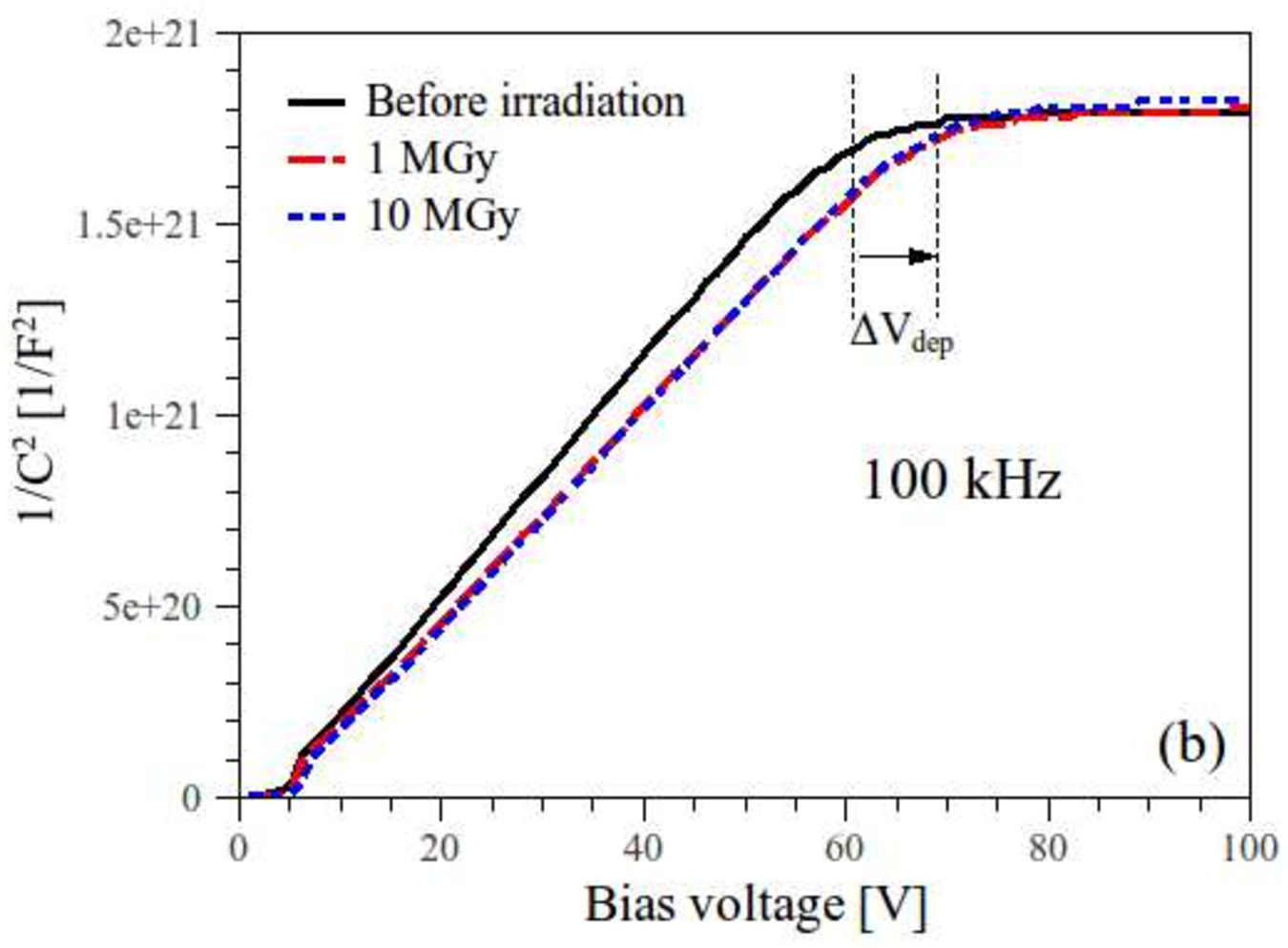}
\includegraphics[scale=0.5]{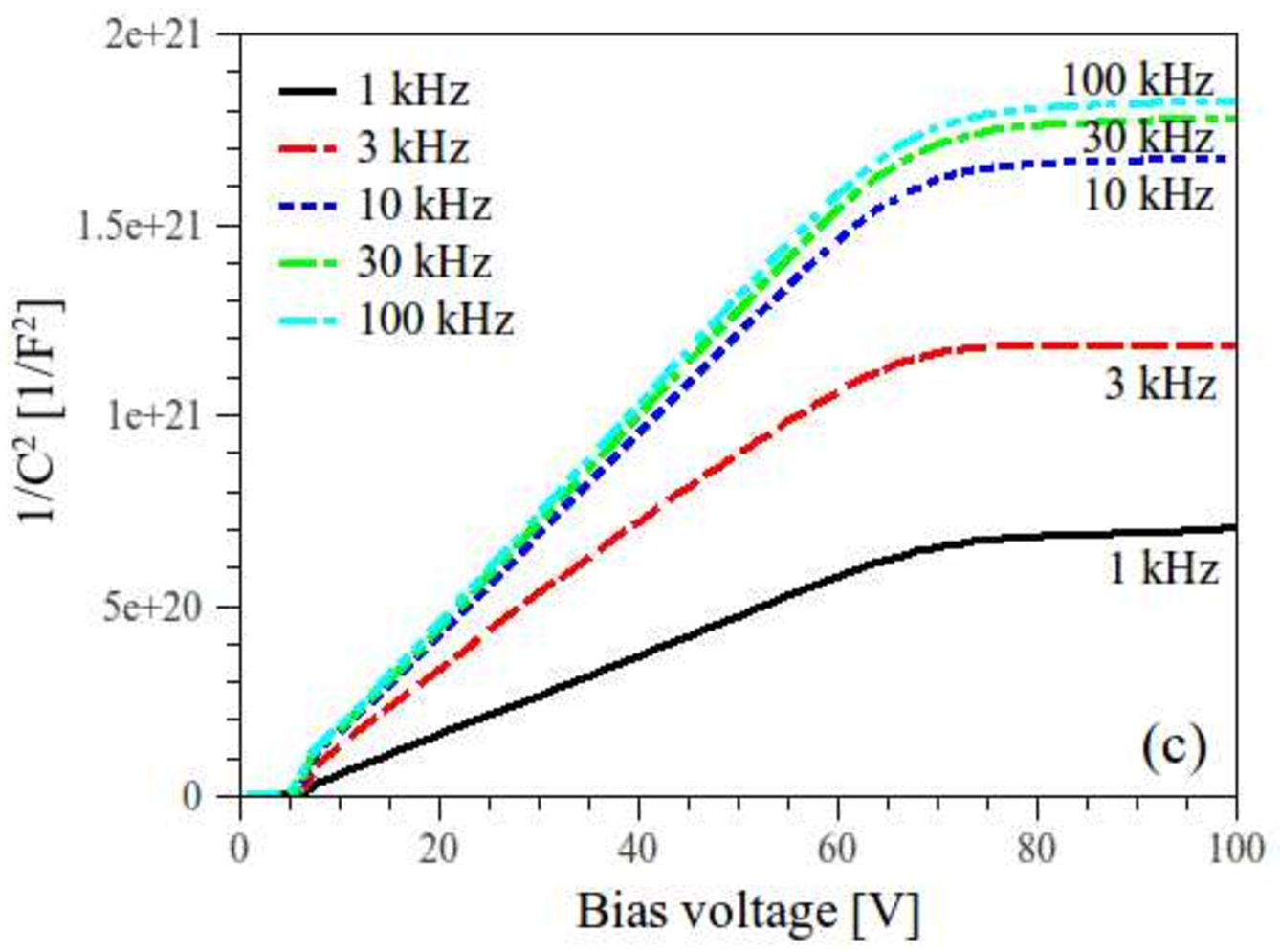}
\includegraphics[scale=0.5]{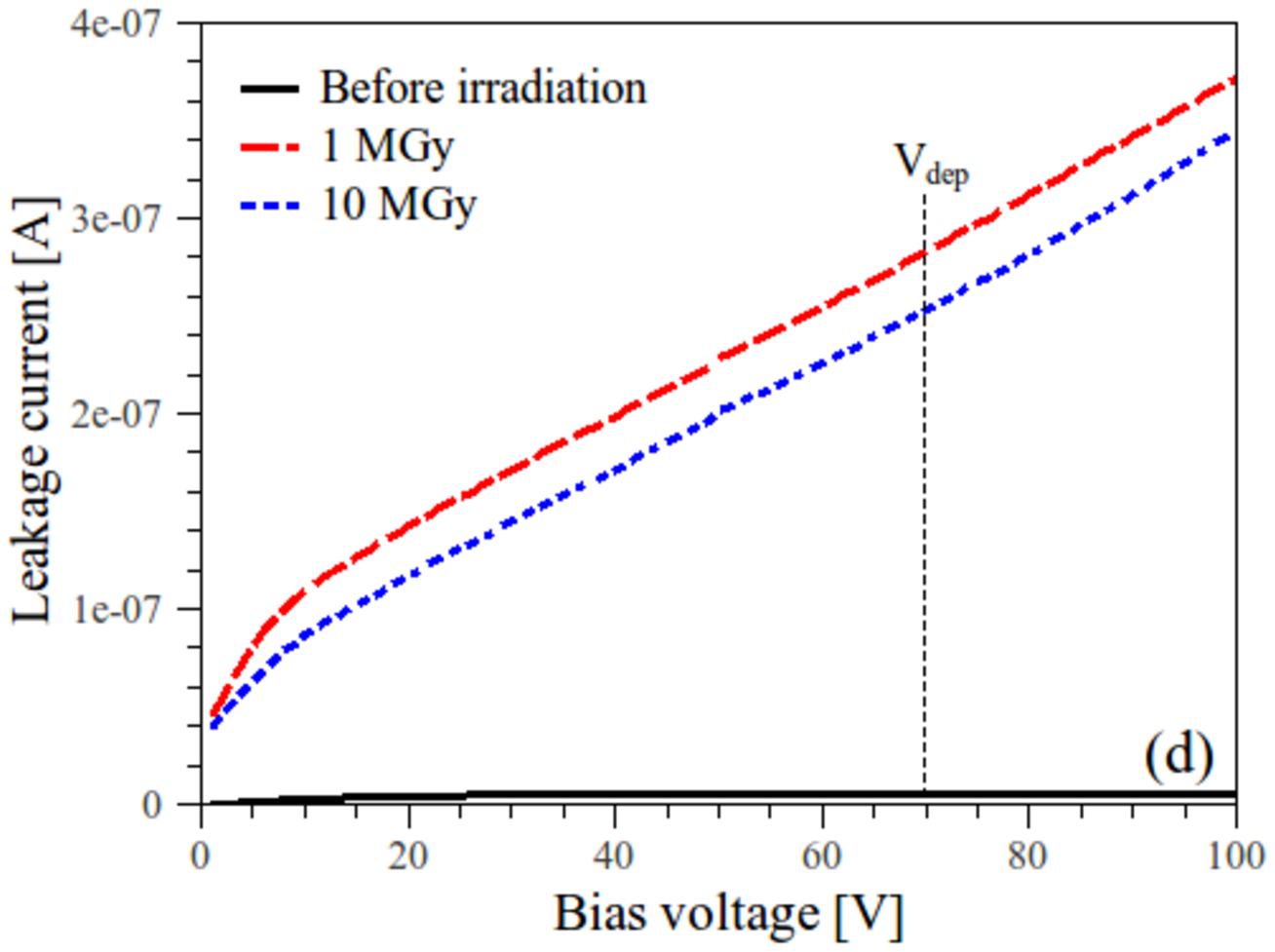}
\caption{(a) I-V characteristics of a single biasing resistor. (b) The capacitance of the entire sensor at 100 kHz, shown as 1/C$^{\text{2}}$ versus bias voltage: Solid - 0 MGy; dashed - 1 MGy; dotted - 10 MGy. The data for 1 and 10 MGy are practically indistinguishable. (c) Sensor capacitance as function of bias voltage for 10 MGy dose. (d) Sensor leakage current: Solid - 0 MGy; dashed - 1 MGy; dotted - 10 MGy.}
\label{Figure5}
\end{figure}

\subsection{Charge losses}

We have studied time resolved current pulses, produced by a sub-nano second laser pulse focused to $\sigma$ = 3 $\mu$m, in the unirradiated and irradiated AC-coupled microstrip sensors as function of the position between the p$^+$ implants. The penetration of the 660 nm laser light in silicon is about 3 $\mu$m. It was found that the measured transient current and the integrated charges collected by the electrodes strongly depend on the irradiation dose of X-rays, biasing history, humidity and time. The shape of the transient signals and the integrated charge as function of the position of the light spot can be described by a model considering losses of either electrons or holes in the "weak" electric field region close to the Si-SiO$_{2}$ interface. A further publication which describes the method, presents the results on charge losses and discusses the relevance for X-ray doses is in preparation.

\section{Summary and outlook}

MOS capacitors and AC-coupled microstrip sensors built on high resistivity <100> n-type silicon have been irradiated with 12 keV X-rays to doses up to 1 GGy and 100 MGy, respectively. For the MOS capacitors results on the fixed oxide charge density and interface trap density as function of dose and the procedure to extract the microscopic radiation damage parameters have been presented. It was found that at least three dominant interface traps have to be used to describe the measurements and to obtain a good description of the C/G-V curves. For doses above 10 - 100 MGy the densities of the oxide charges and of the three interface traps saturate. For the AC-coupled microstrip sensors, it was found that the change of the electrical properties, e.g. capacitance, leakage current and interstrip capacitance and resistance, and charge collection after irradiation are related to the width of the electron accumulation layer formed close to the Si-SiO$_{2}$ interface. Its size increases with dose and decreases with applied bias voltage. It also depends on the biasing history, time and environmental conditions like humidity and temperature. The extracted parameters have been implemented into the Synopsys TCAD simulation program. They provide a reasonable description of the observations on silicon microstrip sensors.

\acknowledgments

        The work was performed in the framework of the project "Radiation Damage" financed by the European XFEL-GmbH in close collaboration with the AGIPD project. J. Zhang would like to thank the Marie Curie Initial Training Network "MC-PAD" for financial support. We thank the crew of the HASYLAB DORIS III beam lines and colleagues within AGIPD collaboration for their help with setting up the irradiation facility and discussing the results. The work was also supported by the Helmholtz Alliance "Physics at the Terascale".

\end{document}